\documentclass[12pt] {article}

\usepackage{amssymb,amsmath}
\usepackage{color}
\usepackage{gastex}

\allowdisplaybreaks[1]
\newcommand{\beq}[1]{  \begin{equation} \label{#1} }  
\newcommand{\eeq}{     \end{equation}}  
\newcommand{\beqa}[1]{ \begin{eqnarray} \label{#1} }	
\newcommand{\eeqa}{\end{eqnarray}  }

\renewcommand{\appendix}{
  \setcounter{section}{0}\renewcommand{\thesection}{\Alph{section}}
  \section*{Appendix} 
}

\newcommand{\rf}[1]{(\ref{#1})}

\def\bd#1{\mbox{\boldmath$\displaystyle\mathbf{#1}$} }

\def\singlespacing{\baselineskip=13pt}

\newtheorem{lem}{Lemma}

\newcommand{\bfq}{{\bf q}}
 \newcommand{\hbS}{\widehat{\bf S}}
 \newcommand{\hS}{\widehat{S}}
\newcommand{\hbI}{\widehat{\bf I}} 
\newcommand{\hbR}{\widehat{\bf R}}  \newcommand{\hbQ}{\widehat{\bf Q}}

\begin{document}
\pagestyle{myheadings} \markright{{\sc Extreme values of Poisson's ratio }  } 
\singlespacing

\title{ \textcolor{blue}
{Extreme values of Poisson's ratio and  other \\ engineering moduli
in anisotropic materials} }

\author{A. N. Norris\\ \\    Mechanical and Aerospace Engineering, 
	Rutgers University, \\ Piscataway NJ 08854-8058, USA \,\, norris@rutgers.edu } 
\maketitle

\begin{abstract} 

Conditions for a maximum or minimum of Poisson's ratio of anisotropic elastic  materials are  derived.    For a uniaxial stress in the $1-$direction and Poisson's ratio $\nu$ defined by the contraction in the  $2-$direction, the following three quantities vanish at a stationary value: $s_{14}$, $[2\nu s_{15} + s_{25}]$ and $[(2 \nu -1)s_{16} + s_{26}]$, where  $s_{IJ}$ are the components of the compliance tensor.  Analogous conditions for stationary values of Young's modulus and the shear modulus are obtained, along with second derivatives of the three engineering moduli at the stationary values.  The stationary conditions and the hessian matrices are presented in forms that are independent of the coordinates, which lead to simple search algorithms for extreme values.  In each case  the global extremes can be found by a simple search over the stretch direction $\bf n$ only.  Simplifications for stretch directions in a plane of orthotropic symmetry are also presented, along with numerical examples  for the extreme values of the three engineering constants in  crystals of monoclinic symmetry. 

\end{abstract}


\section{Introduction}\label{sec1}

Poisson's ratio $\nu$,  Young's modulus $E$ and the shear modulus $G$, collectively called the engineering moduli, are of fixed value in isotropic materials and related by $2G(1+\nu)=E$.  No such connection holds in anisotropic elastic solids, and all they become dependent upon the directions of stretch, lateral strain, and the shear directions.  Hayes \cite{Hayes72} derived  some universal relations between values for certain pairs of orthogonal directions.  However, apart from cubic symmetry \cite{Norris05d} there is no general formula for the directions and values associated with the largest and smallest values of the engineering moduli.  The purpose of this paper is to provide systematic methods which can be used to find the extreme values of the engineering moduli in any type of anisotropy.

The problem of finding the extreme values of Young's modulus is  the simplest   since $E$ depends only on a single direction of stretch.  Numerical searching is practical and  straightforward; thus  Cazzani and Rovati  provide a detailed analysis of the extrema of Young's modulus  for cubic and transversely isotropic materials \cite{Cazzani03} and for materials with tetragonal symmetry \cite{Cazzani05}, with extensive illustrative examples. 
Boulanger and Hayes \cite{bh95} obtained analytic expressions related to extrema of  Young's modulus.  For stretch in the 1-direction,  they showed that    $E=1/s_{11}$ achieves a stationary value if the two conditions $s_{15}=0$ and $s_{16}=0$ are satisfied.  
In a pair of complementary papers, Ting derived explicit expressions for the stress directions and the stationary values  of  Young's modulus  for triclinic and monoclinic \cite{Ting05a}, orthotropic, tetragonal, trigonal, hexagonal and cubic materials \cite{Ting05}.  
 We will rederive the stationary conditions for $E$ below, along with conditions required for a local maximum or minimum. 

Poisson's ratio   and the shear modulus depend upon  pairs of orthogonal directions, which makes their classification far more complicated than for $E$.  At the same time, there is considerable interest in anisotropic materials which exhibit negative values of Poisson's ratio, also called auxetic materials \cite{Yang04}. 
In sharp contrast to isotropic solids for which $-1< \nu < 1/2$, the value of $\nu$ is   unrestricted in anisotropic materials and may achieve arbitrarily large positive and negative values in the same material.  The first hint of this surprising possibility was given by 
Boulanger and Hayes \cite{bh98b} who presented a theoretical set of elastic moduli for a material with orthorhombic symmetry which satisfy  the positivity requirements, but exhibit simultaneous  arbitrarily large positive and negative values of $\nu$.  Ting and Chen \cite{ting2005b} and Ting \cite{ting2005c} subsequently demonstrated that the same remarkable phenomenon can be obtained in any non-isotropic material symmetry, including cubic symmetry and transverse isotropy.    Further explanation  of the effect in cubic symmetry is provided in Section \ref{sec6} below and in \cite{Norris05d}.    We note that  Rovati presented extensive numerical examples of auxetic behavior in orthorhombic  \cite{Rovati03a}  and   monoclinic materials \cite{Rovati04},   while 
Ting and Barnett \cite{ting2005}  derived general conditions required for the occurrence of negative values of  $\nu$. 
 

The purpose of this paper is to provide a general framework for finding the maximum and minimum values of $\nu$ and $G$ in  anisotropic materials.  Some progress in this regard is due to Ting \cite{Ting05} who discusses the conditions for extreme values of the shear modulus with particular attention to shear in planes of material symmetry.   As far as I know, there are no results reported to date on conditions necessary for extreme values of the Poisson's ratio.   Particular attention is given in this paper to the Poisson's ratio, with the emphasis on deriving conditions that are independent of the coordinate system used.  It will become evident that there is a strong analogy between the problems for the shear modulus and the Poisson's ratio.  In particular, by formulating the problems in a coordinate free manner, the task of searching for extreme values of both is similar. 

The outline of the paper is as follows.  
The three engineering moduli  are introduced in Section \ref{sec2}, along with the equations for the transformation of  elastic moduli under rotation. These are used in  Section \ref{sec3} to derive the conditions required for stationary values of $\nu$, $E$ and $G$, and the values of the second derivatives (Hessian) at the stationary points are determined in Section \ref{sec4}.   The various results are all cast in terms of stretch, strain and shearing along coordinate axes.  The more general format for stationary conditions in general directions, independent of the coordinates, are presented in Section \ref{sec5}. The specific application to Poisson's ratio is considered in Section \ref{sec6}.  The stationary conditions for stretch in a plane of orthotropic symmetry are derived, and it is  shown that at most four stationary values of $\nu$  can occur, two for in-plane lateral strain, and two  out-of-plane.  These results are applied to the specific case of extreme values of $\nu$ in materials with cubic symmetry, recovering results of \cite{Norris05d}.  Applications to generally anisotropic materials are also discussed, and a fast procedure for searching for extreme values of $\nu$ is derived and demonstrated for some materials of monoclinic symmetry.  Finally, in Section \ref{sec7} we present a similar procedure for finding  the global  extreme values of $G$ in generally anisotropic media, with numerical examples. 

\section{Definition of the engineering moduli and preliminary equations}\label{sec2}

Poisson's ratio  measures lateral strain in the presence of uniaxial stress.  For any orthonormal pair of vectors $\{ {\bf n}, {\bf m} \}$, the Poisson's ratio $\nu_{nm} = \nu( {\bf n}, {\bf m})  $ is defined by the ratio of the strains in the two directions for a uniaxial state of stress along one of them \cite{Rovati04}: 
\beq{eq1}
\nu_{nm} = -\frac{  \bd{\varepsilon}:{\bf m}{\bf m} }{  \bd{\varepsilon} :{\bf n}{\bf n} }, \qquad \mbox{for } \, 
\bd{\sigma} = \sigma  \, \bf{n}   \bf{n} , 
\eeq 
where $\bd{\varepsilon}$ and $\bd{\sigma}$ are the symmetric tensors of strain and stress, respectively, and 
${\bf a}{\bf b} $ is the tensor product, sometimes denoted ${\bf a}\otimes {\bf b}$.   The Young's modulus $E_n = E({\bf n})$ relates the axial strain and stress, 
\beq{093}
E_n = \frac{ \sigma }{  \bd{\varepsilon} :{\bf n}{\bf n} } \quad \mbox{for } \, 
\bd{\sigma} = \sigma  \, \bf{n}   \bf{n} .
\eeq 
The third engineering modulus is the shear modulus  $G_{nm} = G( {\bf n}, {\bf m})  $,  
\beq{094}
G_{nm} = \frac{ \sigma }{   \bd{\varepsilon} :(\bf{n}   \bf{m}+  \bf{m} \bf{n}) } \quad \mbox{for } \, 
\bd{\sigma} = \sigma  \, (\bf{n}   \bf{m}+  \bf{m} \bf{n}). 
\eeq 
Tensor components are defined relative to the fixed orthonormal basis $\{ {\bf e}_1,  {\bf e}_2,  {\bf e}_3 \}$, 
\beq{d1}
\bd{\sigma} = \sigma_{ij}  \, {\bf e}_i   {\bf e}_j, 
\qquad
{\bd{\varepsilon}} = \varepsilon_{ij} \,  {\bf e}_i   {\bf e}_j. 
\eeq
The  
 stress $\sigma_{ij}$ and strain $\varepsilon_{ij}$ are related by\footnote{Lower case Latin suffices take on the values 1, 2, and 3, and the summation convention on repeated indices is assumed unless noted otherwise.}  
\beq{00}
\varepsilon_{ij} = s_{ijkl}\sigma_{kl}. 
\eeq
 Here $s_{ijkl}$ denote the components of the fourth order compliance tensor.  We  use the  Voigt notation for conciseness;  compliance is  
${\bf S} = \left[ s_{IJ} \right], \, I,J=1,2,\cdots,6$, with $I =1,2,3,4,5, 6$ corresponding to  $ij = 11,22,33,23,31,12$, and $s_{JI} = s_{IJ}$.

The goal is to find conditions for a maximum or minimum of the engineering moduli, with emphasis on Poisson's ratio.
With no loss in generality assume that the $\bd n$ in the ${\bf e}_1$ direction, 
and  $\bd m$ is in the ${\bf e}_2$ direction.  Thus, we consider 
$\nu \equiv  \nu_{12}$, $E \equiv E_1$ and  $G \equiv G_{12}$, i.e.\footnote{Note that we take $s_{66}= s_{1212}$ although it is common to subsume the factor of $4$ in the definition of $s_{66}$ in eq. $\rf{1e1}_3$.} 
\beq{1e1} \nu =   - \frac{s_{12}}{ s_{11}  },
\qquad
E =   \frac1{ s_{11}  },
\qquad
G =  \frac1{ 4s_{66}  }.
\eeq
 Our objective  is then to find conditions for a maximum or minimum of each engineering modulus under the assumption that the material is assumed to be free to orient in arbitrary directions with oriented moduli   while the stress remains of fixed orientation.  This is equivalent to stationarity conditions for $\nu_{nm}$, $E_n$ and  $G_{nm}$ for  a fixed orientation material while  $\{ {\bf n}, {\bf m} \}$ range over all possible orthonormal pairs.  

We therefore need to consider how $\nu$, $E$ and $G$  of \rf{1e1} vary under general  rotation of the material.  Define the rotation by angle $\theta$ about an arbitrary direction ${\bfq}$, $|{\bfq} | = 1$ , as ${\bf Q} ({\bfq} , \theta )\in O(3)$, such that vectors (including the basis vectors) transform as ${\bf r} \rightarrow {\bf r}' = {\bf Q}{\bf r}$. 
Under the change of basis associated with ${\bf Q} ({\bfq} , \theta )$, second order tensors (including stress and strain) transform as $\bd{\sigma} \rightarrow \bd{\sigma} '$, where $\sigma_{ij} ' = Q_{ir}Q_{js}   \sigma_{rs}$, 
or 
\beq{qsat3}
 \sigma_{ij} ' = {\cal Q}_{ijrs}  \sigma_{rs},  
 \qquad \mbox{where}\quad 
{\cal Q}_{ijrs} = \frac12   \big( Q_{ir}Q_{js} + Q_{is}Q_{jr}\big). 
\eeq

In order to simplify the algebra we use the connection  between   fourth order elasticity tensors in 3 dimensions and second order symmetric tensor of 6 dimensions \cite{c3}.  Accordingly, 
the $6\times 6$ matrix   $\hbS$ with elements $  \hS_{IJ}$  is defined as  
\beq{defs}
\hbS = {\bf T}   {\bf S}  {\bf T},
\quad \mbox{where  }
{\bf T} \equiv 
{\rm diag} \big(1,\,1,\,1,\,\sqrt{2},\,\sqrt{2},\,\sqrt{2} \big)   .
\eeq
Rotation of second and fourth order tensors is most simply presented in terms of the $6\times 6$ rotation matrix $\hbQ $  which is the 6-dimensional version of the fourth order tensor ${\cal Q}_{ijrs}$, introduced by Mehrabadi et al. \cite{mcj}. 
Fourth order tensors  transform as $\hbS \rightarrow \hbS ' = \hbQ\hbS\hbQ^T$, where 
$\hbQ  ({\bfq} , \theta )$ is  an orthogonal second order tensor of six 
dimensions, satisfying $\hbQ\hbQ^T = \hbQ^T\hbQ = \hbI = $diag$(1,1,1,1,1,1)$.  It satisfies   
\beq{3.1}    
 \frac{\partial \hbQ}{\partial \theta} ({\bfq} , \theta ) =  \hbR({\bfq}) \hbQ , \qquad  \hbQ({\bfq},0)=\hbI,
 \eeq
 where  $\hbR$ is a skew symmetric six dimensional tensor  linear in  $\bfq$, 
\beq{a1b}
\hbR ({\bfq}) =  \begin{pmatrix}
0 & 0 & 0 & 
        0 & \sqrt{2} q_2 & -\sqrt{2} q_3  
\\ 
0 & 0 & 0 & 
        -\sqrt{2} q_1 &0 &  \sqrt{2} q_3
\\ 
0 & 0 & 0 & 
        \sqrt{2} q_1 &  -\sqrt{2} q_2 &0
\\ 
0 & \sqrt{2} q_1 & -\sqrt{2} q_1   & 0 & q_3 & -q_2
\\ 
-\sqrt{2} q_2 &0 &  \sqrt{2} q_2  & -q_3 & 0 & q_1
\\ 
\sqrt{2} q_3 &  -\sqrt{2} q_3 &0 &q_2 & -q_1 & 0 
\end{pmatrix} . 
\eeq
Further details can be found in \cite{mcj,Norris05}.

\section{Extremal conditions}\label{sec3}

Consider any one of the engineering moduli, say $f$,  as a function of both the underlying compliance and of the rotation $\hbQ$.  A  stationary value is obtained if $f$ is unchanged  with respect to additional small rotations.  In order to formulate this more precisely, 
assume $f$ is at a stationary point, and define 
\beq{s6}
 \hbS ( {\bfq}, \theta) =  \hbQ({\bfq} , \theta ) \hbS \hbQ^T ({\bfq} , \theta ) \, .   
\eeq
Define the rotational derivative, 
\begin{align}\label{s7}
f'( {\bfq} ) &\equiv 
  \left. \frac{\partial f }{\partial \theta }  \big( \hbS ( {\bfq}, \theta) \big)\right|_{\theta = 0} 
  \nonumber \\
   &= \frac{\partial f }{\partial s_{IJ} }\, s_{IJ}'( {\bfq} ).  
\end{align}
The elements $s_{IJ}'( {\bfq} )$ of the 
the rotational derivative of the compliance can be expressed by using the representation \rf{3.1} with \rf{s6}, 
\beq{s8}
 \hbS ' =  \hbR({\bfq}) \hbS   + \hbS  \hbR^T ({\bfq})\, .  
\eeq
Thus, 
\beq{map}
\begin{pmatrix}
   s_{11}'
 \\  s_{22}' 
 \\  s_{33}' 
 \\  s_{12}' 
 \\  s_{23}' 
 \\  s_{13}' 
 \\  s_{14}'
 \\  s_{25}' 
 \\  s_{36}'
 \\  s_{15}'
 \\  s_{16}'
 \\  s_{24}'
 \\  s_{26}'
 \\  s_{34}'
 \\  s_{35}'
 \\  s_{44}'
 \\  s_{55}'
 \\  s_{66}'
 \\  s_{45}'
 \\  s_{46}'
 \\  s_{56}'
 \end{pmatrix}
 = 
 \begin{pmatrix}
 0  & 4s_{15}  &  - 4 s_{16}  \\  
 -4  s_{24}   & 0 &    4 s_{26}  \\  
   4 s_{34}   & - 4 s_{35}  & 0  \\
 -2 s_{14}   &    2 s_{25}  &    2 s_{16} -  2s_{26}   \\  
  2 s_{24} -  2s_{34}   & - 2 s_{25}  &     2 s_{36}  \\   
   2 s_{14}   &   2 s_{35} -  2s_{15}  &  - 2 s_{36}  \\   
  s_{12}  - s_{13}   &  -  s_{16} +2 s_{45}     &   s_{15}- 2 s_{46}  \\ 
 s_{26}-2 s_{45}   & s_{23}  - s_{12}  &  - s_{24}   + 2 s_{56}  \\   
 -s_{35}  + 2 s_{46}   &  s_{34} -2 s_{56}  &   s_{13} - s_{23}  \\  
  s_{16}   & -  s_{11}   + s_{13} + 2 s_{55}  &  - s_{14}    - 2 s_{56}  \\   
   -s_{15}   & s_{14} + 2 s_{56}  &   s_{11} - s_{12}    - 2 s_{66}  \\   
 s_{22}- s_{23} - 2 s_{44}    & - s_{26}  &   s_{25}+2 s_{46}     \\  
 -s_{25} - 2 s_{46}   &  s_{24}  &  -  s_{22}+ s_{12}    + 2 s_{66}  \\  
   - s_{33} + s_{23} + 2 s_{44}   &   - s_{36}   - 2 s_{45}  &     s_{35}  \\ 
  s_{36}  + 2 s_{45}   &  s_{33}  - s_{13}- 2 s_{55}   &  - s_{34}  \\       
   2s_{24} -  2s_{34}   &   - 2 s_{46}  &     2 s_{45}  \\   
  2 s_{56}   &  2 s_{35} - 2s_{15}  &  - 2 s_{45}  \\   
   -2 s_{56}   &  2 s_{46}  &    2 s_{16} -  2s_{26}  \\ 
  s_{25} - s_{35} + s_{46}    & -   s_{14}  + s_{34} - s_{56}  &   s_{55} - s_{44}  \\    
  -s_{36}  + s_{26}   - s_{45}    &   s_{44} - s_{66}   &     s_{14} - s_{24} + s_{56}  \\  
   s_{66} - s_{55}   & s_{36} -s_{16} + s_{45}   &   - s_{25} + s_{15} - s_{46}   
  \end{pmatrix}
  \begin{pmatrix} 
  q_1 \\ \\  q_2 \\ \\  q_3
   \end{pmatrix}.
\eeq
Here  $s_{IJ}$ are the values at $\theta = 0$, which are independent of $\bfq$.  
The derivatives in eq. \rf{map} are linear functions of the coordinates of $\bfq$, and so we may write
\beq{921}
f'( {\bfq} ) = {\bf d}^{(f)}\cdot \bfq , 
\eeq
where the vector $ {\bf d}^{(f)}$ is independent of $\bfq$ and depends only on the compliances.

The engineering modulus $f$ is stationary with respect to the direction $\bf n$, and the 
direction $\bf m$ where applicable, if $f ' ( {\bfq} ) $  vanishes for {\em all} axes of rotation $\bfq$.  This is equivalent to requiring that all possible deviations in  $\bf n$ and $\bf m$ leave $f$ unchanged to first order in the rotation.  The stipulation that this hold for all   rotation axes covers all permissible transformations.   The general condition for stationarity is therefore that  the vector ${\bf d}^{(f)}$ must vanish, i.e., 
\beq{922}
{\bf d}^{(f)} = 0 \quad\mbox{at a stationary point of }f. 
\eeq
We now apply this formalism to the three engineering moduli and derive 
$ {\bf d}^{(\nu)}$, $ {\bf d}^{(E)}$ and $ {\bf d}^{(G)}$ in turn. 

\subsection{Poisson's ratio}
For Poisson's ratio, eq. \rf{s7} becomes 
\beq{1e3}
 \nu ' ( {\bfq} ) = s_{11}^{-2}\, \big[ s_{12}\,  s_{11}'( {\bfq} ) -
s_{11}\,  s_{12}'( {\bfq} )\big].  
\eeq
Thus, from eqs. \rf{map} through \rf{1e3}, we have 
\beq{1e5}
{\bf d}^{(\nu)}= 2s_{11}^{-2}\, \big[
s_{11}  s_{14}  {\bf e}_1 + \big( 2 s_{12}   s_{15}  -s_{11} s_{25} \big) {\bf e}_2 + \big(s_{11} s_{26} - s_{11} s_{16} - 2
s_{12} s_{16}  \big) {\bf e}_3 \big] \, .
 \eeq
Setting this to zero and using the definition of $\nu$ in \rf{1e1}, and the fact that $s_{11} >0$, we obtain  three conditions for a stationary value of Poisson's ratio:  
\begin{subequations} \label{e1}
\begin{align}\label{e1a}
 s_{14} &= 0, \\
 2\nu s_{15} + s_{25}&= 0, \label{e1b} \\
    (2 \nu -1)s_{16} + s_{26} &= 0 \label{e1c}. 
\end{align}
\end{subequations}
These must be simultaneously satisfied at a maximum or minimum of $\nu$. 
Note that the compliance elements appearing in \rf{e1} are all identically zero in isotropic materials.

\subsection{Young's modulus}
Proceeding in the same manner as for the Poisson's ratio,  and using $E ' ( {\bfq} ) = -  s_{11}^{-2} s_{11}'( {\bfq} ) $, gives 
\beq{071}
 {\bf d}^{(E)}
 = 4s_{11}^{-2}\, \big( -s_{15} {\bf e}_2 +  s_{16}  {\bf e}_3 \big) . 
\eeq
Setting this to zero  implies   the conditions for an extremum in Young's modulus  
 \begin{subequations} \label{073}
\begin{align}\label{073a}
 s_{15} &= 0, \\
s_{16} &= 0 . 
\end{align}
\end{subequations}
These agree with   two conditions determined by Boulanger and Hayes \cite{bh95} and by Ting
\cite{Ting05a}. 

\subsection{Shear modulus}
The rotational derivative of the shear modulus is $  G ' ( {\bfq} ) = - \frac14 s_{66}^{-2} s_{66}'( {\bfq} ) $, and  the gradient vector is 
\beq{0701}
 {\bf d}^{(G)}
  = \frac{1}{2} s_{66}^{-2}\, \big[ s_{56} {\bf e}_1 -s_{46} {\bf e}_2 +  (s_{26}- s_{16})  {\bf e}_3 \big]. 
\eeq
Hence, the shear modulus has an extreme value if the following three conditions hold: 
 \begin{subequations} \label{074}
\begin{align}\label{074a}
 s_{56} &= 0, \\
  s_{46} &= 0, \\
s_{16} - s_{26} &= 0 . 
\end{align}
\end{subequations}

\section{Second derivatives}\label{sec4}

The nature of a stationary value of the general engineering modulus $f$ can be discerned, at least locally, by the second derivative.  By analogy with eq. \rf{s7}, we define the rotational second derivative, 
\beq{s71}
f''( {\bfq} ) \equiv 
   \frac{\partial^2 f }{\partial s_{IJ}\partial s_{KL} }\, s_{IJ}'( {\bfq} ) s_{KL}'( {\bfq} )  + \frac{\partial f }{\partial s_{IJ} }\, s_{IJ}''( {\bfq} )   .  
\eeq
The elements $s_{IJ}''( {\bfq} )$ of the 
the rotational second derivative of the compliance 
follow from  
\beq{s8a}
 \hbS '' ({\bfq}) =  \hbR^2 \hbS   + \hbS  \hbR^{2T} + 2\hbR \hbS\hbR^T  \, .  
\eeq
This is a direct consequence of eq. \rf{s8}.  
We do not need all 21 elements, and for brevity only present  the following three values which are necessary to evaluate the second derivatives of the engineering moduli, 
\begin{subequations} \label{2.2}
\begin{align}\label{2.2a}
s_{11}'' &= 4\big[ 
  (s_{13}- s_{11} +2 s_{55} )  q_2^2 +   (s_{12} - s_{11} + 2s_{66} ) q_3^2   - 2(s_{14} + 2 s_{56} ) q_2 q_3  
  +   s_{15}q_3q_1 +   s_{16}q_1q_2   \big] ,\quad 
 \\
 s_{12}'' &= 2 \big[ 
  (s_{13} - s_{12} )q_1^2  +   ( s_{23} - s_{12} )  q_2^2 
 +  (s_{11} + s_{22} -2 s_{12} -4 s_{66}  )  q_3^2 
 \nonumber \\ & \qquad 
 +   ( s_{14} -2s_{24}+ 4 s_{56}  ) q_2 q_3 +   ( s_{25}  -2 s_{15} +4 s_{46} )q_3 q_1  
 +   (s_{16} + s_{26} -4 s_{45}  ) q_1 q_2   \big],  \quad \label{2.2b}
 \\
 s_{66} '' &= 
2 \bigr[   ( s_{55} - s_{66} ) q_1^2 
 +( s_{44} - s_{66} ) q_2^2 
 +( s_{16} + s_{26} - 2  s_{36}- 2  s_{45} ) q_1 q_2 
 \nonumber \\ & \qquad 
 +( s_{11} +  s_{22} - 2  s_{12} - 4  s_{66}  ) q_3^2
 +( 2 s_{25} - 2 s_{15} +   3 s_{46} )  q_1q_3
 \nonumber \\ & \qquad 
 +( 2s_{14} -2 s_{24} +3  s_{56}) q_2 q_3
   \bigr]\, . 
\end{align}
\end{subequations} 

Before applying these to the three engineering moduli $f=\nu$, $E$ and $G$, we note that in each case that $f$ is a homogeneous function of degree 0 or - in the compliance elements. Consequently  the second derivative $f'' $  evaluated at the stationary point where  $f' = 0$ simplifies because  the first term in   \rf{s71} vanishes, leaving
\beq{s73}
f''( {\bfq} ) =
    \frac{\partial f }{\partial s_{IJ} }\, s_{IJ}''( {\bfq} )   
    \qquad \mbox{at  } f'( {\bfq} ) =0.  
\eeq
The terms $ s_{IJ}''( {\bfq} )$ are second order in ${\bfq}$, and  we can write
\beq{s74}
f''( {\bfq} ) = {\bf D}^{(f)}:\bfq \bfq  
\qquad \mbox{at  } f'( {\bfq} ) =0,
\eeq
where ${\bf D}^{(f)} = {{\bf D}^{(f)}}^T$ is a non-dimensional symmetric $3\times 3$ matrix which is independent of ${\bfq}$.  
Thus, ${\bf D}^{(f)}$ is positive (negative) semi-definite at a local minimum (maximum) of $f$.   
The condition for a local minimum (maximum) is therefore that the three eigenvalues of 
${\bf D}^{(f)}$ are positive (negative).  
If the matrix is not definite and has eigenvalues of opposite sign, then the  modulus has a locally saddle shaped behavior.   

\subsection{Poisson's ratio}
The second derivative of the Poisson's ratio at a 
 stationary point is, using \rf{s73}, 
\beq{2e2}
 \nu '' ( {\bfq} ) =  - s_{11}^{-1}\,  \big( s_{12}'' +\nu  s_{11} ''\big)   
 \quad \mbox{at } \, \nu ' ( {\bfq} ) = 0.
\eeq
Thus, when $\nu ' ( {\bfq} ) = 0$, eqs.  \rf{2.2} and \rf{2e2}  give 
\begin{align}\label{2.5}
 \nu '' ( {\bfq} ) &=  \frac{2}{s_{11}^2}\,  \bigg\{
 s_{11}( s_{12}-  s_{13} )  q_1^2 
 +  \big[ s_{11}(  s_{12}-  s_{13}) + 2s_{12}(s_{13}-  s_{11}+2 s_{55} )\big] q_2^2
 \nonumber \\
 & \qquad + \big[ s_{11}(  2s_{12}+4s_{66} -  s_{11}-  s_{22})+ 2s_{12}(s_{12}-  s_{11}+2 s_{66} )\big] q_3^2
 \nonumber \\
 & \qquad + 2\big[ s_{11}(  s_{24}-2s_{56} -  \frac12 s_{14}) - 2s_{12}(s_{14}+2 s_{56} )\big]q_2q_3 
 \nonumber \\
 & \qquad +  2\big[ s_{11}(  s_{15}-2s_{46} -  \frac12 s_{25}) +s_{12}s_{15}\big]q_3q_1
  \nonumber \\
 &  \qquad +  2\big[ s_{11}(  2s_{45} -  \frac12 s_{16}-  \frac12 s_{26}) +s_{12}s_{16}\big]q_1q_2 \bigg\} \, . 
\end{align}
Since this is evaluated at the stationary value, we may use \rf{e1}  to simplify and obtain
\beq{2.8}
 {\bf D}^{(\nu)}=  \frac{2}{s_{11}}       
 \begin{pmatrix}
 s_{12}-  s_{13}  &   2s_{45} -   s_{16} &   s_{15}-2s_{46} 
 \\ && \\ 
   & s_{12}-  s_{13} -    &  
  \\ 2s_{45} -   s_{16} & \ \  2\nu (s_{13}-  s_{11}+2 s_{55} ) & s_{24}-2(1-2\nu )s_{56} 
  \\ && \\ 
  & & s_{12}-  s_{22}+2s_{66} + \\ 
  s_{15}-2s_{46}  &    s_{24}-2(1-2\nu )s_{56}  & \ \ (1- 2\nu )(s_{12}-  s_{11}+2 s_{66} )  
\end{pmatrix}, 
\eeq
where we have used the definition of $\nu$ to simplify the elements. 

\subsection{Young's modulus}
At a stationary point  
\beq{81}
E '' ( {\bfq} ) =  - s_{11}^{-2}\,  s_{11} ''
 \quad \mbox{at } \, E ' ( {\bfq} ) = 0.
\eeq
Using eq. \rf{2.2} and the extremal conditions \rf{073}, we find 
\beq{75}
  {\bf D}^{(E)} =  \frac{1}{s_{11}^2}       
 \begin{pmatrix}
 0 & 0& 0 \\
 0 &  s_{11}- s_{13}-  2s_{55} &   s_{14}+2 s_{56}
 \\ 
  0 &  s_{14}+2 s_{56} & s_{11}- s_{12}-  2s_{66}
\end{pmatrix}.
\eeq
Note that $  {\bf D}^{(E)}$ is rank deficient (of rank 2), which is a consequence of the fact that $E$ is invariant under rotation about the ${\bf e}_1$ stretch axis.   The local nature of the stationary value depends upon the two non-zero eigenvalues of the matrix.

\subsection{Shear modulus}
The shear modulus satisfies  
\beq{84}
G '' ( {\bfq} ) =  - \frac14 s_{66}^{-2}\,  s_{66} ''
 \quad \mbox{at } \, G ' ( {\bfq} ) = 0, 
\eeq
and hence, 
\beq{87}
  {\bf D}^{(G)}=  \frac{1}{2s_{66}^2}       
 \begin{pmatrix}
s_{66} - s_{55}  &  s_{36} - s_{16} +  s_{45} &    s_{15} -  s_{25}
 \\ 
s_{36} - s_{16} +  s_{45}  &  s_{66} - s_{44}  &   s_{24} - s_{14}
  \\ 
 s_{15} -  s_{25} &  s_{24} - s_{14}   &  2  s_{12} + 4  s_{66} -s_{11} -  s_{22}
 \end{pmatrix}.
\eeq

\section{Coordinate invariant formulation}\label{sec5}

In this section we rephrase the results for the stationary conditions and for the second derivatives at the stationary conditions in coordinate invariant form.  
Let $\{ {\bf n}, {\bf m}, {\bf p}\}$ be an orthonormal triad analogous to $\{ {\bf e}_1, {\bf e}_2, {\bf e}_3\}$ before.  Define the non-dimensional  symmetric second order tensors ${\bf A}, {\bf B},{\bf N},{\bf M}$ and ${\bf P}$, as follows:
\begin{subequations} \label{5.1}
\begin{align}\label{5.1a}
&A_{ij} = s_{nn}^{-1}s_{ijkl}n_kn_l,\qquad
B_{ij} = s_{nn}^{-1}s_{ijkl}m_km_l, \qquad
C_{ij} = s_{nn}^{-1}s_{ijkl}p_kp_l, 
\\ 
&N_{ij} = s_{nn}^{-1}s_{ikjl}n_kn_l, \qquad
M_{ij} = s_{nn}^{-1}s_{ikjl}m_km_l, \qquad
P_{ij} = s_{nn}^{-1}s_{ikjl}p_kp_l.   
\end{align}
\end{subequations}
where $s_{nn}= s_{ijkl}n_in_jn_kn_l$.   Thus, 
\beq{5.2}
\nu({\bf n},{\bf m}) = - {\bf A}:{\bf m}{\bf m},
\qquad 
E({\bf n})  = 1/s_{nn}, 
\qquad
G({\bf n},{\bf m}) = E({\bf n})/\big( 4{\bf N}:{\bf m}{\bf m}\big). 
\eeq

\subsection{Poisson's ratio}

The derivative of Poisson's ratio can  be expressed in general form as 
\beq{5.4}
 {\bf d}^{(\nu )} = 
 \big[ {\bf A}:({\bf m}{\bf p}+{\bf p}{\bf m})\big] {\bf n}
 - \big[ (2\nu {\bf A}+ {\bf B}):({\bf p}{\bf n}+{\bf n}{\bf p})\big] {\bf m}
 + \big\{ \big[(2\nu -1) {\bf A}+ {\bf B}\big]:({\bf n}{\bf m}+{\bf m}{\bf n})\big\} {\bf p}\, ,
 \qquad
 \eeq
which follows from eq. \rf{1e5}, and may be checked by the substitution 
$\{ {\bf n}, {\bf m}, {\bf p}\}$ $\rightarrow$ $\{ {\bf e}_1, {\bf e}_2, {\bf e}_3\}$.  This provides a local expansion of the Poisson's ratio for small rotation about the axis $\bfq$: 
\beq{5.5}
\nu  ( {\bfq} , \theta) = \nu  ( {\bfq} , 0) +  {\bf d}^{(\nu )}\cdot \bfq \, \theta + {\rm O}(\theta^2)\, .
 \eeq
In particular, $ {\bf d}^{(\nu )} = 0$  at a stationary point. 
  
The second order tensor ${\bf D}^{(\nu)}$ of \rf{2.8} becomes, in general format, 
\begin{align}\label{5.6}
{\bf D}^{(\nu)} &= \big[ {\bf A}:({\bf m}{\bf m}-{\bf p}{\bf p})\big]{\bf n}{\bf n}
+ \big\{\nu  -[ (1+2\nu) {\bf A}+4 \nu  {\bf N}]:{\bf p}{\bf p}
 \big\}{\bf m}{\bf m}
	\nonumber \\ &
\quad 	+ \big\{2\nu^2-1  +[ 4(1-\nu) {\bf N}- {\bf M}]:{\bf m}{\bf m}
 \big\}{\bf p}{\bf p}
	\nonumber \\ &
\quad  + \big[ (2{\bf P}- {\bf N}):({\bf n}{\bf m}+{\bf m}{\bf n})\big]
\frac12	({\bf n}{\bf m}+{\bf m}{\bf n})
	\nonumber \\ & 	
\quad 	+ \big[ ({\bf N}- 2{\bf M}):({\bf p}{\bf n}+{\bf n}{\bf p})\big]\frac12	 
		({\bf p}{\bf n}+{\bf n}{\bf p})
	\nonumber \\ &	
\quad  + \big\{ \big[{\bf M} - 2(1-2\nu) {\bf N}\big]:({\bf m}{\bf p}+{\bf p}{\bf m})\big\}
\frac12		({\bf m}{\bf p}+{\bf p}{\bf m})\, . 
\end{align}
Hence, the local expansion near a stationary point is 
\beq{5.7}
\nu  ( {\bfq} , \theta) = \nu  ( {\bfq} , 0) + \frac12 {\bf D}^{(\nu)}: \bfq \bfq\, \theta^2 + {\rm O}(\theta^3)\, .
\eeq

\subsection{Young's modulus}
In the same way as before  we find that 
\beq{5.41}
 {\bf d}^{(E )} = 2E( {\bf n})\, 
 \big[ - {\bf A}:({\bf p}{\bf n}+{\bf n}{\bf p})\,  {\bf m}
 + {\bf B}:({\bf n}{\bf m}+{\bf m}{\bf n})\, {\bf p}\big]  , 
 \eeq
 and
\begin{align}\label{5.42}
 {\bf D}^{(E )} = E( {\bf n})\, \big\{ 
 &
 \big[ {\bf A}:({\bf n}{\bf n}-{\bf p}{\bf p}) - 2{\bf P}:{\bf n}{\bf n}\big]\, {\bf m}{\bf m}
  + \big[ {\bf A}:({\bf n}{\bf n}-{\bf m}{\bf m}) - 2{\bf M}:{\bf n}{\bf n}\big]\, {\bf p}{\bf p}
  \nonumber \\   
  & +  ({\bf A} +2 {\bf N}):({\bf m}{\bf p}+{\bf p}{\bf m})\big\}
\frac12		({\bf m}{\bf p}+{\bf p}{\bf m})
 \big\} . 
\end{align}

\subsection{Shear modulus}
Similarly, for the shear modulus
\beq{5.43}
 {\bf d}^{(G )} = 4\frac{G^2( {\bf n})}{E( {\bf n})}\, 
 \big[ {\bf N}:({\bf m}{\bf p}+{\bf p}{\bf m})\,  {\bf n} 
 - {\bf M}:({\bf p}{\bf n}+{\bf n}{\bf p})\,  {\bf m}
 + ({\bf B}-{\bf A}):({\bf n}{\bf m}+{\bf m}{\bf n})\, {\bf p}\big]  , 
 \eeq
 and
\begin{align}\label{5.44}
 {\bf D}^{(G )} =& 8\frac{G^2( {\bf n})}{E( {\bf n})}\, \big\{ 
  {\bf N}:({\bf m}{\bf m}-{\bf p}{\bf p})\, {\bf n}{\bf n}
  + {\bf M}:({\bf n}{\bf n}-{\bf p}{\bf p})\, {\bf m}{\bf m}  
  	+ 4{\bf N}:{\bf m}{\bf m}\, {\bf p}{\bf p}
  \nonumber \\   
 & +  ({\bf P} +{\bf C}- {\bf A}):({\bf n}{\bf m}+{\bf m}{\bf n}) 
\frac12		({\bf n}{\bf m}+{\bf m}{\bf n})
 +  ({\bf A} - {\bf B}):\big[  ({\bf m}{\bf m}-{\bf n}{\bf n}) \, {\bf p}{\bf p}  
 \nonumber \\   
 & -({\bf m}{\bf p}+{\bf p}{\bf m}) 
\frac12		({\bf m}{\bf p}+{\bf p}{\bf m})
 + -({\bf p}{\bf n}+{\bf n}{\bf p}) 
\frac12		({\bf p}{\bf n}+{\bf n}{\bf p}) \big]   \big\} . 
\end{align}

\section{Applications to Poisson's ratio}\label{sec6}

We now concentrate on general properties of the Poisson's ratio, applying the formalism for the stationary value to different situations.  We begin by consider the general case of a plane of material symmetry in an orthotropic material. 

\subsection{Plane of symmetry in orthotropic material}
We assume the stretch direction ${\bf n} $ lies in a plane of symmetry of an orthotropic material, and the direction of contraction ${\bf m}$ lies (a) perpendicular to the plane or (b)  in the plane.
This configuration includes all planes  in hexagonal materials that contain the axis of symmetry, and therefore provides the stationary values of $\nu$ in materials with hexagonal symmetry (transverse isotropy).    

With no loss in generality, let ${\bf e}_3^{(0)}$ be the normal to the plane of symmetry and let  case (a) correspond to $ {\bf m} ={\bf e}_3$ and case (b) corresponds to 
$ {\bf m} $ in the plane of ${\bf e}_1^{(0)},{\bf e}_2^{(0)}  $.  In both (a) and (b) $ {\bf n} $ lies in the plane of ${\bf e}_1^{(0)},{\bf e}_2^{(0)}$.   
Define the rotated axes 
\beq{051}
 {\bf e}_1= \cos\theta \, {\bf e}_1^{(0)} + \sin\theta \, {\bf e}_2^{(0)},
\qquad
{\bf e}_2 = -\sin\theta  \, {\bf e}_1^{(0)} + \cos\theta \, {\bf e}_2^{(0)}, 
\qquad
{\bf e}_3 ={\bf e}_3^{(0)} .  
\eeq
Let $S_{IJ}$ denote the compliances relative to the fixed set of axes $\{ {\bf e}_1^{(0)}, {\bf e}_2^{(0)}, {\bf e}_3^{(0)} \}$, and $s_{IJ}$ the compliances in the  coordinates of the rotated axes.  By definition of a symmetry plane, all coefficients $s_{ijkl}$ with index $3$ appearing once or thrice are zero.  Then, 
\begin{subequations}\label{734}
\begin{align}\label{734a}
s_{11}& =  
\frac{S_{11}S_{22} - S_0^2}{S_{11}+S_{22} - 2S_0}
+ \frac 14 (S_{11}+S_{22} - 2S_0)\big[ 
\frac{S_{22} - S_{11}}{S_{11}+S_{22} - 2S_0} 
- \cos2 \theta\big]^2 , 
\\
s_{12} &=  S_{12} + \frac 14 (S_{11}+S_{22} - 2S_0)\, \sin^2 2 \theta ,
 \\
 s_{13} &=  \frac12 (S_{13} +S_{23}) - \frac12 (S_{23} -S_{13})\, \cos 2 \theta ,
 \\
s_{16} &=  \frac 14 (S_{11}+S_{22} - 2S_0)\big[ 
 \frac{S_{22} - S_{11}}{S_{11}+S_{22} - 2S_0} 
- \cos 2 \theta\big]\, \sin 2 \theta,\label{734b}
 \\
s_{26} &=  \frac 14 (S_{11}+S_{22} - 2S_0)\big[ 
 \frac{S_{22} - S_{11}}{S_{11}+S_{22} - 2S_0} 
+ \cos 2 \theta\big]\, \sin 2 \theta,
\\
 s_{36} &=  \frac12 (S_{23} -S_{13})\, \sin 2 \theta  ,  \label{734c}
\end{align}
\end{subequations}
 where
\beq{733}
S_0 \equiv S_{12}+2S_{66}\, . 
\eeq
Thus, in the two cases to be considered, we have $ {\bf n}= {\bf n}_1$, and   $s_{11} = 1/E(\theta)$.

\subsubsection*{(a) ${\bf m} = {\bf e}_3$ perpendicular to  the plane of symmetry}

We   now consider  stationary values of $\nu = \nu_{13}$, for which the three conditions for stationary $\nu$ are, instead of \rf{e1}, 
\beq{042}
 s_{14} = 0, \qquad
 2\nu s_{16} + s_{36}= 0, \qquad 
    (2 \nu -1)s_{15} + s_{35} = 0 \, . 
\eeq
The first and last of these are automatically satisfied, based on the assumed material symmetry. 
Using $s_{16}$ and $s_{36}$ from \rf{734}, the second of \rf{042} implies that $\sin 2 \theta = 0$, which is the exceptional case of prior symmetry, or that $\theta$ satisfies
\beq{752}
\cos 2 \theta = \frac{S_{22} - S_{11}}{S_{11}+S_{22} - 2S_0} 
+ \frac1{\nu} \big(\frac{S_{23} - S_{13}}{S_{11}+S_{22} - 2S_0}\big)\, . 
\eeq
Using this to  eliminate $\cos 2 \theta$ from the expression for  $\nu = \nu_{13}= -  s_{13}/ s_{11}$
yields a quadratic equation for $\nu$, 
\beq{046}
\nu^2 - \nu_a \nu - \frac14 {\rho_a} =0  ,  
\eeq
where
\beq{402}
\nu_a = \frac{(S_0- S_{22} ) S_{13} + (S_0 - S_{11} )S_{23}}{S_{11}S_{22} - S_0^2},
\qquad
\rho_a= \frac{(S_{23} - S_{13})^2}{S_{11}S_{22} - S_0^2}. 
\eeq

Define $E^*$ and $\theta^*$ by 
\beq{343}
E^* = \frac{S_{11}+S_{22} - 2S_0}{S_{11}S_{22} - S_0^2}, 
 \eeq
then we may identify $E^*$ as the value of $E(\theta)$ when the second term on the RHS of \rf{734a} vanishes, i.e. $E^* =E(\theta^*)$, where $\theta^*$ satisfies 
\beq{763}
\cos2 \theta^* = 
\frac{S_{22} - S_{11}}{S_{11}+S_{22} - 2S_0}  . 
\eeq
The angle $\theta^*$ defines the direction at which $E$ is stationary (maximum or minimum).  It exists iff the RHS of \rf{763} lies between $-1$ and $1$.  Regardless of whether or not the angle exists, it can be checked, $\nu_a = \nu_{13}(\theta^*) = -E^* s_{13}(\theta^*)$.   The value of Young's modulus in the stretch direction $\theta$ for the stationary value of $\nu$  satisfies
\beq{6412}
\frac{E^*}{E} + \frac{\nu_a}{\nu } = 2 . 
\eeq

In summary, the possible stationary values for stretch in the plane of symmetry and the strain measured in the   direction perpendicular to the plane are 
\beq{834}
\nu_{a\pm} = \frac12 {\nu_a} \pm \frac12 \big( \nu_a^2 + \rho_a)^{1/2}, 
\eeq
and the stationary values occur if $-1< \gamma_{a\pm} < 1$, where 
\beq{7521}
\gamma_{a\pm} = \frac{S_{22} - S_{11}}{S_{11}+S_{22} - 2S_0} 
+ \frac1{\nu_{a\pm}} \big(\frac{S_{23} - S_{13}}{S_{11}+S_{22} - 2S_0}\big)\, ,  
\eeq
in which case the direction of stretch is given by $\theta = \frac12 \cos^{-1}\gamma_{a\pm}$.  Otherwise the stationary values occur at $ \theta = 0$ and $\pi/2$. 

\subsubsection*{(b) ${\bf m}= {\bf e}_2$ in the plane of symmetry}

With ${\bf n} = {\bf e}_1$ again,  conditions 
\rf{e1a} and \rf{e1b} are met and only \rf{e1c} is not automatically satisfied.  
Substituting the   expressions for $s_{16}$ and $s_{26}$ from \rf{734} into equation \rf{e1c} implies that either $\sin 2 \theta = 0$, which is simply the axial case, or 
\beq{751}
\cos 2 \theta = \frac{S_{22} - S_{11}}{S_{11}+S_{22} - 2S_0} + \frac{1}{\nu-1}\, \big(\frac{S_{22} - S_{11}}{S_{11}+S_{22} - 2S_0}\big)\, . 
\eeq
Substituting this into the relation for $\nu = \nu_{12}$: $\nu s_{nn} + s_{12} = 0$ and eliminating $\cos 2 \theta$ produces a quadratic equation in $\nu$.  The equation is most simply expressed as a quadratic in the shifted Poisson's ratio $(\nu - 1)$: 
\beq{764}
(\nu - 1)^2 - (\nu_b - 1)(\nu - 1)- \frac14 {\rho_b} = 0, 
\eeq
where
\beq{4022}
\nu_b = -1 -E^*\, \big( S_{12} - S_0\big),
\qquad
\rho_b= \frac{(S_{11} - S_{22})^2}{S_{11}S_{22} - S_0^2} . 
\eeq
That is, $\nu_b = \nu_{12}(\theta^*) = -E^* s_{12}(\theta^*)$.
Note that the value of Young's modulus $E$ in the stretch direction $\theta$ satisfies
\beq{641}
\frac{E^*}{E} + \frac{\nu_b - 1}{\nu - 1} = 2 . 
\eeq

In summary, the possible stationary values for stretch and strain both in the plane of symmetry  are 
\beq{8341}
\nu_{b\pm} = \frac12 (\nu_b +1) \pm \frac12 \big( (\nu_b-1)^2 + \rho_b)^{1/2}, 
\eeq
and the stationary values occur if $-1< \gamma_{b\pm} < 1$, where 
\beq{7522}
\gamma_{b\pm} = \frac{S_{22} - S_{11}}{S_{11}+S_{22} - 2S_0} 
+ \frac1{\nu_{b\pm}-1} \big(\frac{S_{22} - S_{11}}{S_{11}+S_{22} - 2S_0} \big)\, ,  
\eeq
in which case the direction of stretch is given by $\theta = \frac12 \cos^{-1}\gamma_{b\pm}$. 

\subsection{Example: cubic materials}

The general coordinate invariant form of the compliance of a cubic material is \cite{walpole84}
\beq{sc2}
{\bf S} =\big(  \frac{1}{3\kappa} - \frac{1}{2\mu_2}\big) \, {\bf J} + \frac{1}{2\mu_1} \,  {\bf I}
+ \big( \frac{1}{2\mu_2} - \frac{1}{2\mu_1} \big)\,  {\bf D} , 
\eeq
where ${\bf I}$ and ${\bf J}$ are fourth order isotropic tensors,  
$I_{ijkl} = \frac12 (\delta_{ik}\delta_{jl}+ \delta_{il}\delta_{jk})$, 
$J_{ijkl} = \frac13 \delta_{ij}\delta_{kl}$, and 
\beq{defd}
{\bf D} = {\bf a}{\bf a}{\bf a}{\bf a}+
{\bf b}{\bf b}{\bf b}{\bf b}+
{\bf c}{\bf c}{\bf c}{\bf c} \, . 
\eeq
Here, the orthonormal triad $\{ {\bf a},\, {\bf b},\, {\bf c}\}$ is coaxial with the cube axes.  The condition that the elastic strain energy is always positive definite is that the three moduli $\kappa$, $\mu_1$ and $\mu_2$ are positive.   We use, for simplicity,   crystallographic-type notation for unit vectors, e.g.,  $ pq\bar{r}$ where $p$, $q$ and $r$ are positive numbers, indicates the unit vector  $(p{\bf a} +q{\bf b} -r{\bf c} )/\sqrt{p^2+q^2+r^2}$. 

\bigskip
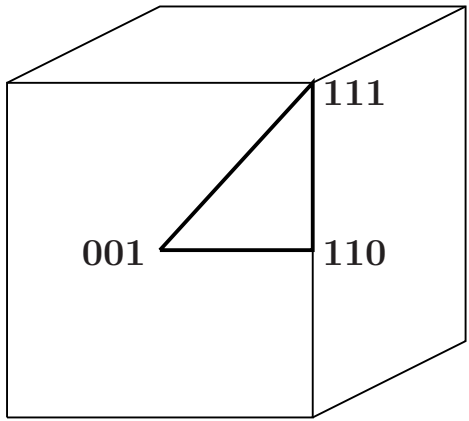
\begin{figure}[htbp]
				\begin{center}	
 	
\setlength{\unitlength}{.2in}
\begin{picture}(10,10)(0,0)
\drawline[AHnb=0,linewidth=.05](0,0)(8,0)(8,8.764)(0,8.764)(0,0)
\drawline[AHnb=0,linewidth=.05](0,8.764)(4,10.764)(12,10.764)(12,2)(8,0)
\drawline[AHnb=0,linewidth=.05](8,8.764)(12,10.764)
\drawline[AHnb=0,linewidth=.1](4,4.382)(8,8.764)(8,4.382)(4,4.382)
\put(2.8,4.3){\makebox(0,0){\bf{\large{001}}}}
\put(9.1,4.3){\makebox(0,0){\bf{\large{110}}}}
\put(9.1,8.5){\makebox(0,0){\bf{\large{111}}}}

\end{picture}
	\caption{\small The irreducible $1/48 $th of the cube surface is defined by the isosceles triangle with vertices as shown.  }
		\label{fcube} \end{center}  
	\end{figure}

We consider the stationary conditions \rf{e1} for stretch in the ${\bf e}_1$ direction and lateral strain in the ${\bf e}_2$ direction.  Since $s_{14}$, $s_{15}$, $s_{16}$, $s_{25}$ and  $s_{26}$ vanish for isotropic media, it follows that the only contribution to these quantities is from the tensor  ${\bf D}$.  We may  therefore rewrite the stationary conditions \rf{e1} as
\begin{subequations} \label{cub1}
\begin{align}\label{cub1a}
 D_{14} &= 0, \\
 2\nu D_{15} + D_{25}&= 0, \label{cub1b} \\
    (2 \nu -1)D_{16} - D_{26} &= 0 \label{cub1c},  
\end{align}
\end{subequations}
where $D_{14} = D_{1123}$, etc.   The realm of stretch directions that needs to be considered may be reduced to those defining the irreducible $1/48$th of the surface of the unit sphere. This in turn is defined by $1/48$th of the surface of the cube, see Fig. \ref{fcube}, where the vertices of the triangle correspond to the directions $001$, $110$, and $111$.  In a separate paper \cite{Norris05d} it is shown that the extreme values of $\nu$ do not occur within the interior of the triangle.  It turns out that the extreme values are only possible for stretch direction ${\bf e}_1$ along $001$, $110$, or in certain cases, for ${\bf e}_1$ located along the edge between $110$ and $111$.  In the case that ${\bf e}_1 = 001$, the lateral direction ${\bf e}_2$ may be any orthogonal direction, and when ${\bf e}_1 = 110$ the lateral directions are $001$ or $1\bar{1}0$, each of which can correspond to the minimum or maximum for $\nu$, depending on the elastic parameters $\kappa$, $\mu_1$ and $\mu_2$.  It is clear from the symmetry of the situation that the quantities $D_{14}$, $D_{15}$, $D_{16}$, $D_{25}$ and  $D_{26}$ vanish identically for ${\bf e}_1$ along $001$ or $110$ with ${\bf e}_2$ as described.  A full  description of the possible extreme values of $\nu$ in cubic materials is  involved but complete, and we refer to \cite{Norris05d} for details. 

To summarize the findings of Norris \cite{Norris05d} regarding solutions of eqs. \rf{cub1}: all  possible stretch directions which solve the three stationary conditions are confined to the symmetry plane with normal $110$, and equivalent planes of symmetry.   We can therefore apply the results for a plane of orthotropic symmetry.  Explicit calculation from \rf{sc2} gives 
$S_{66} =  (4\mu_1)^{-1}$, 
$S_{12} = S_{23} = (9\kappa)^{-1} - (6\mu_2)^{-1}$, 
$S_{13} =S_{12} + \chi /4$, 
$S_0 = S_{12} + 2S_{66}$, 
$S_{11} = S_0+\chi /4$, $S_{22} = S_0+\chi /2$, where 
$\chi = (\mu_2)^{-1}- (\mu_1)^{-1}$. 
Thus, 
\beq{533}
\nu_a = \nu_b = \nu_{111} \equiv \frac{3\kappa - 2\mu_1}{6\kappa + 2\mu_1},
\qquad
\rho_a = \rho_b = \rho \equiv \frac16 ( \nu_{111} +1) \big( \frac{\mu_1}{\mu_2}-1\big),
\eeq
where $\nu_{111} = \nu (111, {\bf m})$ is the Poisson's ratio for stretch in the $111$ direction, and is independent of the lateral direction $\bf m$. 

  The actual values of the possible extrema for $\nu$ can be obtained from  equations and \rf{834}, \rf{8341} and \rf{533}.  Skipping over the unedifying details, see \cite{Norris05d}, it can be shown that only the  stationary values   $\nu_{a-}$ and $\nu_{b+}$   are possible global extrema, 
\begin{subequations}
\begin{align}\label{cub10}
\nu_{a-} &= \frac12 \nu_{111} - \frac12 \sqrt{ \nu_{111}^2 + \rho},
\\
\nu_{b+} &= \frac12 \big( \nu_{111}+1\big) + \frac12 \sqrt{ \big( \nu_{111}-1\big)^2 + \rho},
\end{align}
\end{subequations}
Note that $\nu_{111}$  is independent of $\mu_2$, which only enters these expressions via the term $\rho$. 
 The extremely large values of Poisson's ratio discovered by Ting and Chen \cite{ting2005b} correspond  to  $\rho \gg 1$, which can occur if  $\mu_2/\mu_1 \ll 1$.  Under this circumstance  $\nu_{a-}$ is large and negative,  
$\nu_{b+}$ is large and positive, and the magnitudes are, in principle, unbounded \cite{ting2005b,ting2005,Norris05d}.  

The directions associated with the global extrema are given by 
\beq{423}
\cos 2 \theta_{a-} = \frac13 - \frac1{3\nu_{a-}},
\qquad
\cos 2 \theta_{b+} = \frac13 + \frac1{3(\nu_{b+} -1)}.
\eeq
Both directions bifurcate from $\theta = 0$ \cite{Norris05d}, and therefore 
these extreme values   only occur if  $\nu < -\frac12$ or $\nu > \frac32$, respectively.   A complete description of the extrema for all possible values of the elastic moduli is given by Norris \cite{Norris05d}.

\subsection{Application to generally anisotropic materials}

We first present a result which suggests a simple algorithm for searching for global extreme values of $\nu$ in generally anisotropic materials.  

\subsubsection{A local min-max result for Poisson's ratio}

The tensor of second derivatives of $\nu ({\bf n}, {\bf m})$ at the stationary point, ${\bf D}^{(\nu )}$, must be  positive or negative definite in order that the stationary point is a minimum or a maximum, respectively. Consider a possible minimum, then a necessary although not sufficient condition is that the three diagonal elements of ${\bf D}^{(\nu )}$ are positive.  In particular, eq. \rf{5.6} gives  
\beq{p1}
{\bf D}^{(\nu )}:{\bf n}{\bf n} = \nu_{np} - \nu_{nm}
\ge  0.
\eeq
This implies that  $ \nu = \nu_{nm} $ must be strictly less than $ \nu_{np} $. At the same time, the stationary condition ${\bf d}^{(\nu )}=0$  must hold, and in particular, 
\beq{p3}
{\bf d}^{(\nu )}\cdot{\bf n} = 0 \quad
\Rightarrow \quad 
{\bd{\varepsilon}}: {\bf m}{\bf p}= 0 \quad \mbox{for } \, 
\bd{\sigma} = \sigma  \, \bf{n}   \bf{n} . 
\eeq
This implies that the shear strain $\varepsilon_{mp}$ in the ${\bf m},{\bf p}$-plane is zero. 
Hence, for any unit vector ${\bf r}\perp {\bf n} $, we have   
\beq{p3a}
{\bd{\varepsilon}}: {\bf r}{\bf r}= ( {\bf r}\cdot{\bf m})^2\, \varepsilon_{mm}
+ ( {\bf r}\cdot{\bf p})^2\, \varepsilon_{pp}, 
\eeq
or
\begin{align}\label{p4}
 \nu_{nr}  &=  ( {\bf r}\cdot{\bf m})^2\,  \nu_{nm}+ ( {\bf r}\cdot{\bf p})^2\,  \nu_{np} 
 \nonumber \\
 &=    \nu_{nm}+ ( {\bf r}\cdot{\bf p})^2\,  \big( \nu_{np} -  \nu_{nm}\big) 
 \nonumber \\
 & \ge  \nu_{nm} ,
\end{align}
with equality only for ${\bf r} = {\bf m}$.  Thus, we have: \\

\begin{lem} \label{lem1} 
 \, If  $\nu_{nm}$ is a minimum (maximum) value, then it is also a minimum (maximum)  among all possible $\nu_{nr}$ for ${\bf r}$ in the plane perpendicular to $\bf n$.   
\end{lem}

This result  is a direct consequence of the general expression for ${\bf D}^{(\nu )}$.   It implies that if we can satisfy \rf{p3} then the values of $\nu_{nm}$ and $\nu_{np}$ are the extreme values for the given stretch direction $\bf n$.   We next show how this single condition can be achieved. 

\subsubsection{Satisfaction of one extremal condition}

The stationary values of Poisson's ratio occur, in general, for  stretch directions at which 
the vector ${\bf d}^{(\nu )} $ of eq. \rf{5.4} vanishes.   We now show that one of the three components can be made to vanish;  specifically, ${\bf d}^{(\nu )} \cdot {\bf n}   ={\bf A}:({\bf m}{\bf p}+{\bf p}{\bf m}) $ is zero 
for an  appropriate choice of the orthogonal directions  ${\bf m}$ and ${\bf p}$.   
  
We use the fact that the appropriate pair ${\bf m}$ and ${\bf p}$ correspond to stationary values of  ${\bf A}:{\bf m}{\bf m}$ and ${\bf A}:{\bf p}{\bf p} $. These satisfy 
 ${\bf A}:{\bf m}{\bf m}+ {\bf A}:{\bf p}{\bf p} = $tr${\bf A} - 1$ so a maximum in one implies a minimum for the other.   In order to find these directions for a given $\bf n$, consider the function of $\bf m$: 
\beq{1003}
g(\bf m)\equiv 
{\bf A}:{\bf m}{\bf m} - \lambda {\bf m} \cdot {\bf m} - 2\alpha {\bf m} \cdot {\bf n} \, .  
\eeq
Setting to zero the gradient with respect to $\bf m$ implies that ${\bf m}$ satisfies
\beq{1004b}
 {\bf m} =\alpha \big( {\bf A} - \lambda {\bf I} \big)^{-1} {\bf n} \, .  
\eeq
The scalars $\lambda$ and $\alpha$ follow by requiring that ${\bf m} \cdot {\bf n} =0$ and 
${\bf m} \cdot {\bf m} =1$, respectively. The former  implies that  $\lambda$ satisfies
\beq{1005}
{\bf n} \cdot \big( {\bf A} - \lambda {\bf I} \big)^{-1} {\bf n} = 0 \, .  
\eeq
This condition can be rewritten by expanding the inverse in terms of the cofactor matrix of $\big( {\bf A} - \lambda {\bf I} \big)$,  and using the property  ${\bf A}:{\bf n}{\bf n} = 1$, 
which  yields a quadratic in $\lambda$:
\beq{1007}
 \lambda^2 +   \lambda \big(1- {\rm tr}{\bf A}  \big) + {\rm adj}({\bf A}):{\bf n}{\bf n}\, 
= 0 \, .  
\eeq
Here ${\rm adj}({\bf A})$ is the adjoint matrix. If ${\bf A}$ is invertible, then 
${\rm adj}({\bf A}) = ({\rm det}{\bf A}) {\bf A}^{-1}$, and more generally 
the adjoint is the transpose of the cofactor matrix.  
Thus for any type of anisotropy, and for any direction $\bf n$, it is a straightforward to determine the appropriate orthogonal pair $\bf m$, $\bf p$, which automatically give
${\bf d}^{(\nu )} \cdot {\bf n}   =0$. 
Finding the right axes requires solving the quadratic \rf{1007} for $\lambda$.  The associated values of Poisson's ratio are the maximum and minimum for the given direction $\bf n$. In this way, finding extremal values of Poisson's ratio for all possible $\bf n$ is reduced to seeking values which satisfy the remaining two conditions: 
\beq{1009}
 (2\nu {\bf A}+ {\bf B}):{\bf n}{\bf p}=0,\quad
\big[(2\nu -1) {\bf A}+ {\bf B}\big]:{\bf n}{\bf m}=0, 
\eeq
or equivalently, $ 2\nu s_{15} + s_{25}= 0$ and $ (2 \nu -1)s_{16} + s_{26} =0$, respectively.
This is the strategy used to determine the global extrema of $\nu$ in materials with cubic symmetry  \cite{Norris05d}. 

\subsubsection{Algorithm for finding global  extreme values of $\nu$}

Rather than searching for directions $\bf n$ which satisfy the three stationary conditions on $\nu$, Lemma \ref{lem1} suggests that a simple search for maximum and minimum values of Poisson's ratio can be effected as follows.  For a given $\bf n$ define the pair $\nu_\pm ({\bf n})$ by 
\begin{subequations}
\begin{align}\label{47}
\nu_\pm ({\bf n})&= - {\bf A}: {\bf m}_\pm^{(A)}{\bf m}_\pm^{(A)}, 
\\
 {\bf m}_\pm^{(A)} &=\|\big( {\bf A} - \lambda_\pm^{(A)} {\bf I} \big)^{-1} {\bf n}\|^{-1}\,  \big( {\bf A} - \lambda_\pm^{(A)} {\bf I} \big)^{-1} {\bf n},
 \\
  \lambda_\pm^{(A)} &= \frac12 \big( {\rm tr}{\bf A} -1 \big) \pm 
  \frac12 \big[\big( {\rm tr}{\bf A} -1 \big)^2
  - 4\, {\rm adj}({\bf A}):{\bf n}{\bf n}\big]^{1/2} . 
\end{align}
\end{subequations}
The search for global extrema is then a matter of finding the largest and smallest values 
of $\nu_\pm ({\bf n})$ by searching over all possible directions $\bf n$.   In practice, even for triclinic materials with no symmetry, the search only has to be performed over half of the unit sphere $\|{\bf n}\| = 1$, such as  ${\bf n}\cdot {\bf e}\ge 0$ for some fixed direction ${\bf e}$.   In this way, the numerical search is equivalent in complexity to that of finding the global extrema of  Young's modulus.

\begin{table} [t]		\label{t2}
\begin{center}
\medskip 
\begin{tabular}{lcc ccc} \vspace{-0.1in} &&&& &   \\  
\hline  
Material	&	quantity	&	~ value ~  & $n_1$  & $n_2$  & $n_3$   
\\ 
\hline 
\\
Cesium dihydrogen  &	$E_{max}$ &  19.3  & 0.44  &  0.76 &  0.47 \\
\, \, phosphate& $E_{min}$ &     0.53  &  0.99 &   0.00  &  0.06 \\
\, \, (CsH$_2$PO$_4$) & $G_{max}$ &    12.2   &  0.42  &   0.71 &   -0.56  \\
& $G_{min}$ &    0.20  &  0.68 &   0.00 &    0.73  \\
& $\nu_{max}$ &     2.70  &   -0.23 &   0.82 &   0.52 \\
& $\nu_{min}$ &    -1.93  &   -0.49 &    0.85 &   -0.20 \\
\\
Lanthanum niobate 	&	
	$E_{max}$  &   154.8  &  0.45  &  0.87 &  -0.21   \\
\, \, \,  (LaNbO$_4$)&  $E_{min}$ &  2.27  &  -0.50  &  0.00  &   0.87   \\
& $G_{max}$ &  72.57 & -0.28   & 0.71  &  -0.64  \\
& $G_{min}$ &    0.70 &   -0.96&    0.00&    0.27   \\
& $\nu_{max}$ &    3.95 &   0.00  &   1.00  &  0.00    \\
& $\nu_{min}$ &  -3.01 &  0.00  &  1.00  &   0.00 \\  
\\ 
\hline 		 
\end{tabular}
\caption{Extreme values of $E$, $G$ and $\nu$ for two materials of monoclinic symmetry with symmetry plane $n_2=0$.  Units of $E$ and $G$ are GPa.  
Cesium dihydrogen phosphate: $s_{11}$ = 1820.0, $s_{22}$ = 103.0,
$s_{33}$ = 772.0, $s_{44}$ = 33.25, $s_{55}$ = 112.5, $s_{66}$ = 29.25, $s_{12}$ = -219.0,
$s_{13}$ = -1170.0, $s_{23}$ = 138.0, $s_{15}$ = 124.5, $s_{25}$ = -75.0, $s_{35}$ = -90.5,
$s_{46}$ = 8.25.   
Lanthanum niobate: $s_{11}$ = 66.8, $s_{22}$ = 14.8,
$s_{33}$ = 146.0, $s_{44}$ = 5.7, $s_{55}$ = 265.0, $s_{66}$ = 4.675, $s_{12}$ = 16.9, $s_{13}$ = -94.8,
$s_{23}$ = -30.8, $s_{15}$ = 118.0, $s_{25}$ = 45.6, $s_{35}$ = -186.5, $s_{46}$ = 0.95.
Units in (TPa)$^{-1}$ (Data from Landolt \& Bornstein \cite{Landolt})
   }  
\end{center}
\end{table}

This algorithm was applied to data for two crystals of monoclinic symmetry: Cesium dihydrogen phosphate and	Lanthanum niobate, with the results  in Table 1.  The numerical results were obtained by using a 100$\times$100 mesh for the hemisphere $\|{\bf n}\|=1$, ${\bf n}\cdot
{\bf e}_2 \ge 0$. It was found that the extreme values of the engineering moduli are not sensitive to the mesh size, although the values of the directions do change slightly.

\section{Applications to the shear modulus}\label{sec7}

If we compare the  vector derivative function for the shear modulus, ${\bf d}^{(G )}$ of \rf{5.43}, with ${\bf d}^{(\nu )}$ of \rf{5.4}, we note that the components in the $\bf n$ direction have a similar form, but with different matrices involved.  Thus, it is ${\bf N}$ for  
${\bf d}^{(G )}\cdot {\bf n} $, while for ${\bf d}^{(\nu )}\cdot {\bf n}$ 
it is ${\bf A}$.  The properties of the second derivatives 
${\bf D}^{(G )}: {\bf n}{\bf n} $ and ${\bf D}^{(\nu )}: {\bf n}{\bf n} $ are similarly  related.    Proceeding with the same arguments as for the Poisson's ratio, we may deduce

\begin{lem} \label{lem2} 
 \, If  $G_{nm}$ is a minimum (maximum) value, then it is also a minimum (maximum)  among all possible $G_{nr}$ for ${\bf r}$ in the plane perpendicular to $\bf n$.   
\end{lem}
This in turn leads to a similar method for finding the global extrema of $G$.

\subsection{Algorithm for finding global  extreme values of $G$}

Define  the pair $G_\pm ({\bf n})$ by 
\begin{subequations}
\begin{align}\label{48}
G_\pm ({\bf n})&=  E({\bf n}) /\big( {\bf N}: {\bf m}_\pm^{(N)}{\bf m}_\pm^{(N)}\big), 
\\
 {\bf m}_\pm^{(N)} &=\|\big( {\bf N} - \lambda_\pm^{(N)} {\bf I} \big)^{-1} {\bf n}\|^{-1}\,  \big( {\bf N} - \lambda_\pm^{(N)} {\bf I} \big)^{-1} {\bf n},
 \\
  \lambda_\pm^{(N)} &= \frac12 \big( {\rm tr}{\bf N} -1 \big) \pm 
  \frac12 \big[\big( {\rm tr}{\bf N} -1 \big)^2
  - 4\, {\rm adj}({\bf N}):{\bf n}{\bf n}\big]^{1/2} . 
\end{align}
\end{subequations}
Then the global extreme values of $G$  can be found by searching over $\bf n$ only.  Thus, the problem of finding the global extrema of the shear modulus is reduced to the same level of complexity as searching for the maximum and minimum Young's modulus. 
Table 1 summarizes the extreme values of $G$ found using this 
algorithm for two crystals of monoclinic symmetry. 

\section{Conclusions}

 The results of this paper provide a consistent framework for determining the extreme values of the three engineering moduli in materials of any crystal symmetry or none.    
General conditions have been derived which must be satisfied at stationary values of $\nu$, $E$ and $G$.  These are  equations \rf{e1}, \rf{073} and \rf{074}, which have also been cast in forms that are independent of the coordinates used,  in equations \rf{5.4}, \rf{5.41} and \rf{5.43}, respectively.   The associated three hessian matrices   which determine the local nature of the stationary value, maximum, minimum or saddle, are given in equations \rf{5.6}, \rf{5.42} and \rf{5.44}.   The stationary conditions for Poisson's ratio simplify for stretch in a plane of orthotropic symmetry, for which there are at most 4 stationary values of $\nu$.  Two can occur for in-plane stretch and strain, and the other two for out-of-plane strain.  This implies that transversely isotropic materials have at most four stationary values of Poisson's ratio.  The results for the plane of symmetry also reproduce known results for cubic materials  \cite{Norris05d}.   The hessian matrices for $\nu$ and 
$G$ lead to  algorithms for finding the extreme values.  The key is to remove the dependence on  the $\bf m$ direction by explicit representation of the maximum and minimum for a given $\bf n$ direction.  The algorithms have been demonstrated by application to materials of low symmetry.

				
\end{document}